\documentclass[twocolumn]{aastex631}
\usepackage{graphicx}
\usepackage{txfonts}
\usepackage{booktabs, comment}
\usepackage{natbib,twoopt,hyperref}

\accepted{13 October, 2025}

\begin{document}

\title{Survival of the accretion disk in LMC Recurrent Nova 1968-12a: UV--X-ray case study of the 2024 eruption}

\shorttitle{2024 eruption of LMCN 1968-12a in UV and X-ray}
\shortauthors{Basu et al.}

\correspondingauthor{Judhajeet Basu}
\email{judhajeet20@gmail.com}

\author[0000-0001-7570-545X]{Judhajeet Basu}
\affiliation{Indian Institute of Astrophysics, II Block, Koramangala, Bengaluru 560034, India}
\affiliation{Pondicherry University, R.V. Nagar, Kalapet, Pondicherry 605014, India}

\author[0000-0003-3533-7183]{G.C. Anupama}
\affiliation{Indian Institute of Astrophysics, II Block, Koramangala, Bengaluru 560034, India}

\author[0000-0003-0440-7193]{Jan-Uwe Ness}
\affiliation{European Space Agency, European Space Astronomy Centre, E-28692 Madrid, Spain}

\author[0000-0001-6952-3887]{Kulinder Pal Singh}
\affiliation{Department of Physical Sciences, Indian Institute of Science Education and Research Mohali, 140306, Punjab, India}

\author[0000-0002-3927-5402]{Sudhanshu Barway}
\affiliation{Indian Institute of Astrophysics, II Block, Koramangala, Bengaluru 560034, India}

\author[0009-0000-5909-293X]{Shatakshi Chamoli}
\affiliation{Indian Institute of Astrophysics, II Block, Koramangala, Bengaluru 560034, India}
\affiliation{Pondicherry University, R.V. Nagar, Kalapet, Pondicherry 605014, India}

\begin{abstract}
We report on UV and X-ray observations of the 2024 eruption of the recurrent nova LMCN 1968-12a, a rapidly recurring extragalactic system with a $\sim$4.3 year recurrence period and a massive white dwarf (WD). The eruption was discovered on 2024 August 1.8 by \textit{Swift}, and subsequently monitored using \textit{AstroSat}’s UVIT and SXT, along with Swift/UVOT and XRT. The multi-wavelength light curves reveal a rapid UV-optical decline, followed by a plateau phase exhibiting 1.26-day modulations consistent with the orbital period. The Supersoft (SSS) X-ray emission, that emerged by day 5, exhibited a double peak, suggesting variable obscuration that could be due to an inhomogeneous nova ejecta or due to a nova super-remnant along the line of sight. Time-resolved X-ray spectroscopy shows a blackbody component with T $\approx 10^6$ K. The SEDs obtained concurrently in the UV, peaking at T $\approx$ 20,000 K and with a source radius $\sim$2–3 R$_\odot$, are inconsistent with emission from the secondary star or nova photosphere alone. Instead, the UV emission is attributed to an irradiated accretion disk that survived the eruption. The persistent UV plateau and its temperature suggest that the accretion disk was not completely disrupted and resumed activity within days, consistent with recent findings in other rapidly recurring novae such as U~Sco and M31N~2008-12a.

\end{abstract}


\section{Introduction} 
\label{sec:intro}

Stars are often found in binary systems. When such a binary consists of a white dwarf (WD) and a main-sequence or a red/sub-giant star, it can evolve into a close-binary cataclysmic variable (CV) system \citep{Kraft_1964}. Under these circumstances, the WD can accrete matter from the secondary star via a Roche-lobe overflow or winds from the secondary, forming an accretion disk. As the base of the accretion disk becomes denser and hotter (nearing Fermi temperature), the critical limits of thermonuclear runaway (TNR) are reached (\citealt{Townsley_2004, Starrfield_2016} and references therein). What follows is the release of energy ($ \approx 10^{45}$ ergs) and matter (at $v \gtrapprox 500$~km~s$^{-1}$) from the system, known as a nova eruption \citep{Starrfield_2016}. These transients are important sites for studying various physical processes such as TNRs, shock interactions, mass loss mechanisms, and galactic chemical enrichment \citep{Starrfield_2020,Chomiuk_2021}. 

While it is known that an accretion disk is present at late times in old novae \citep{Selvelli_2013,basu_2024a, Gloria_2025}, there is no consensus on the level of damage to the accretion disk during the outburst. It might remain intact in part (or completely, though this is less likely) within the optically thick ejecta. As the photosphere recedes towards the WD and the ejected material becomes transparent, the (surviving) disk may become visible once more. Accretion resumes and the cycle of nova eruptions repeats after a time interval that depends on the WD mass, the accretion rate, and the nature of the secondary \citep{Shara_2018, Hillman_2020}. Systems that have exhibited multiple nova eruptions within timescales of decades or less are classified as recurrent novae (RNe), that is, novae with more than one observed outburst from the same system. In contrast, systems with much longer recurrence timescales, such that only a single eruption has been observed, are termed classical novae (CNe). 

High WD mass and accretion rate drive rapid RNe. Earlier models by \cite{Starrfield_1975, Starrfield_2016} and \cite{Kato_2014} predicted these systems required massive WDs. \cite{Shara_2018} built upon and validated these predictions, confirming massive WDs for RNe. The shorter the recurrence period, the more massive the WD is. Hence, such systems that contain a CO WD (thus not too rich in $\alpha$ elements) are also top contenders for Type Ia supernova progenitors \citep{Hillman_2016, Wang_2018}.

LMC recurrent nova 1968-12a (hereafter LMCN 1968-12a) was the first extragalactic RN to be discovered in 1968. It has erupted several times since then (see Table~\ref{tab:eruption_info}). Recent eruptions reveal a recurrence period of roughly 4.3 years. This is shorter than the recurrence period for any galactic recurrent nova and longer than only two other extragalactic novae, M31N~2008-12a \citep{Darnley_2016, Henze_2018,basu_2024} and M31N~2017-01e \citep{Shafter_2022,Shafter_2024, Chamoli_2025}. It falls in the category of rapidly recurring systems that have massive WDs \citep{Darnley_2021}. Previous studies of past eruptions of this system concluded that the WD in this binary is $>$ 1.3 M$_\odot$, has an orbital period of 1.26 days, and the mass of the ejecta was comparable to that of the accreted mass \citep{Kuin_2020}. 

The most recent outburst of LMCN 1968-12a was discovered on 2024~August~1.8 UT by the \textit{Swift} observatory. The last quiescent detection was on 2024~Aug~1.19 UT in the $UVW1$ filter at 18.19 $\pm$ 0.09 mag \citep{Darnley_2024}. The eruption time, estimated to be the midpoint of these two times, is given in Table~\ref{tab:eruption_info}. This has been used as the reference time throughout this work.

In this paper, we present the UV and X-ray observations of the 2024 outburst based on data obtained with {\it AstroSat} UVIT and SXT, combined with data from {\it Swift} UVOT and XRT. 

\begin{table*}[t]
    \centering    
    \caption{List of all previous known eruptions of LMCN 1968-12a}
    \begin{tabular}{ccccccc}
    \toprule
     \multicolumn{3}{c}{Discovery} & Eruption date & Time since last eruption & SSS $\mathrm{t_{off}}$ date & References \\
     (UT) & mag    & filter & (MJD)  & (days)    & (days) & \\
     \cmidrule(lr){1-3} \cmidrule(lr){4-6} \cmidrule(lr){7-7}
     1968 Dec 16.5 & 10.9 & $\mathrm{m_{pg}}$ & 40206.0 $\pm$ 1.5 & $\cdots$ & $\cdots$ & (1) \\ 
     $\vdots$ & \multicolumn{6}{c}{} \\
     1990 Feb 14.1 & 11.2 & $\mathrm{m_{pv}}$ & 47936.1 $\pm$ ?	 & $\cdots$ & $\cdots$ & (2) \\
     $\vdots$ & \multicolumn{6}{c}{} \\
     2002 Oct 10   & 11.15 $\pm$ 0.02 & V     & 52557.3 $\pm$ 1.0 & $\cdots$ & $\cdots$ & (3)-(4) \\
     $\vdots$ & \multicolumn{6}{c}{} \\
     2010 Nov 21.2 & 11.7 $\pm$ 0.3   & I     & 55521.2 $\pm$ 1.0 & $\cdots$ & $\cdots$ & (5) \\
     2016 Jan 21.2 & 11.5 $\pm$ 0.2   & I     & 57407.4 $\pm$ 0.8 & 1886.2 $\pm$ 1.8 & 56.1 & (4)\\
     2020 May 06.7 & 13.99 $\pm$ 0.08 & g     & 58971.9 $\pm$ 0.4 & 1564.5 $\pm$ 1.2 & 60   & (6)-(7) \footnote{\href{https://www.aavso.org/lmc-v1341-new-recurrent-nova-eruption}{https://www.aavso.org/lmc-v1341-new-recurrent-nova-eruption}} \\
     2024 Aug 01.8 & 9.9 $\pm$ 0.1   & uvm2   & 60523.5 $\pm$ 0.3 & 1551.6 $\pm$ 0.7 & $>$ 50 & (8), this work  \\

    \bottomrule
    \end{tabular}
    \label{tab:eruption_info}

\textbf{Notes:} 
References - (1) \cite{Sievers_1970}, (2) \cite{Liller_1990}, (3) \cite{Pojmanski_2002}, (4)\cite{Kuin_2020}, (5) \cite{mroz_2014}, (6) \cite{Page_2020}, (7) \cite{Schwarz_2020}, (8) \cite{Darnley_2024}

\end{table*}

\section{Observation and data}
\label{section:obs}

\begin{figure*}
    \centering
    \includegraphics[width=0.49\linewidth]{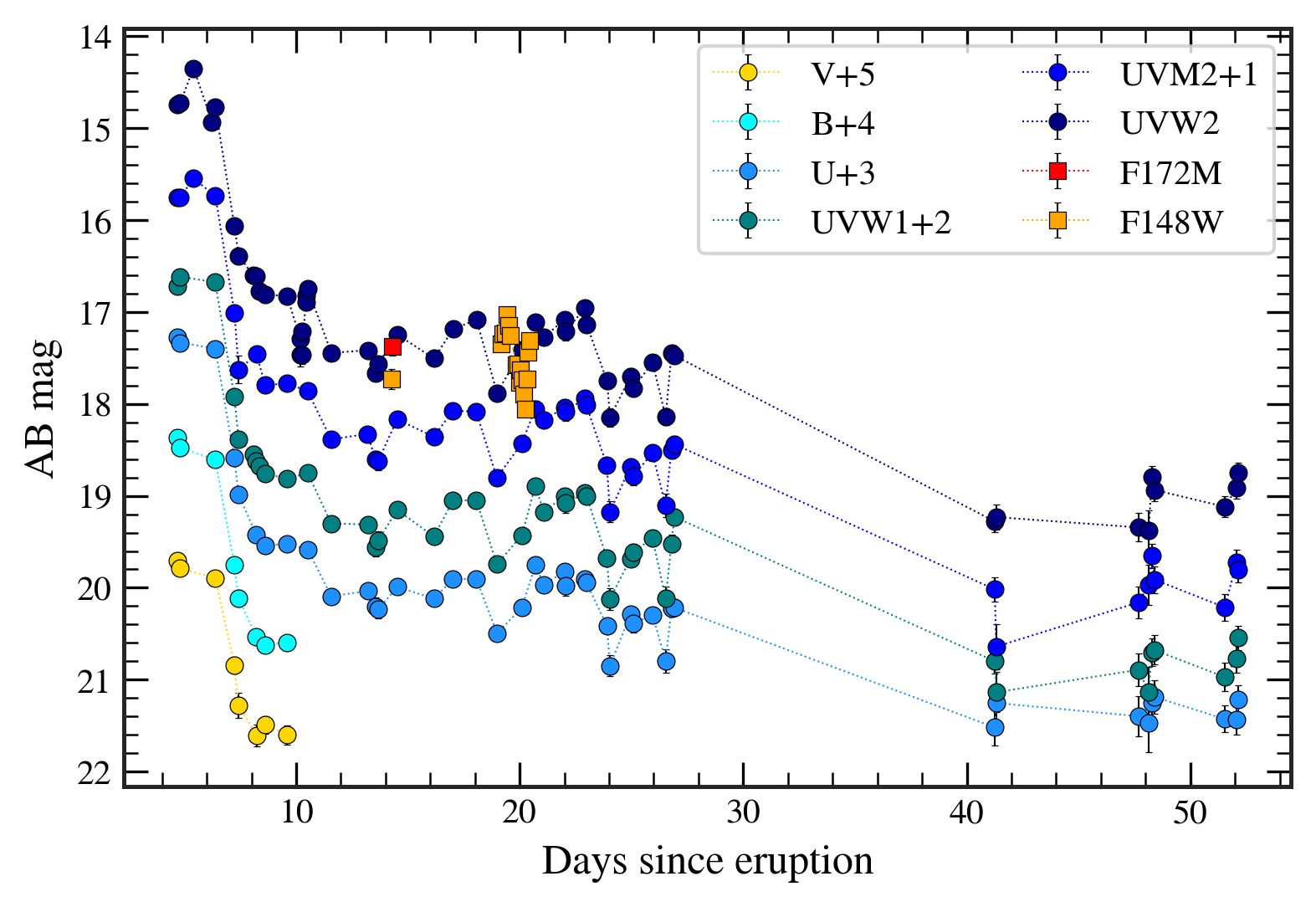}
    \includegraphics[width=0.49\linewidth]{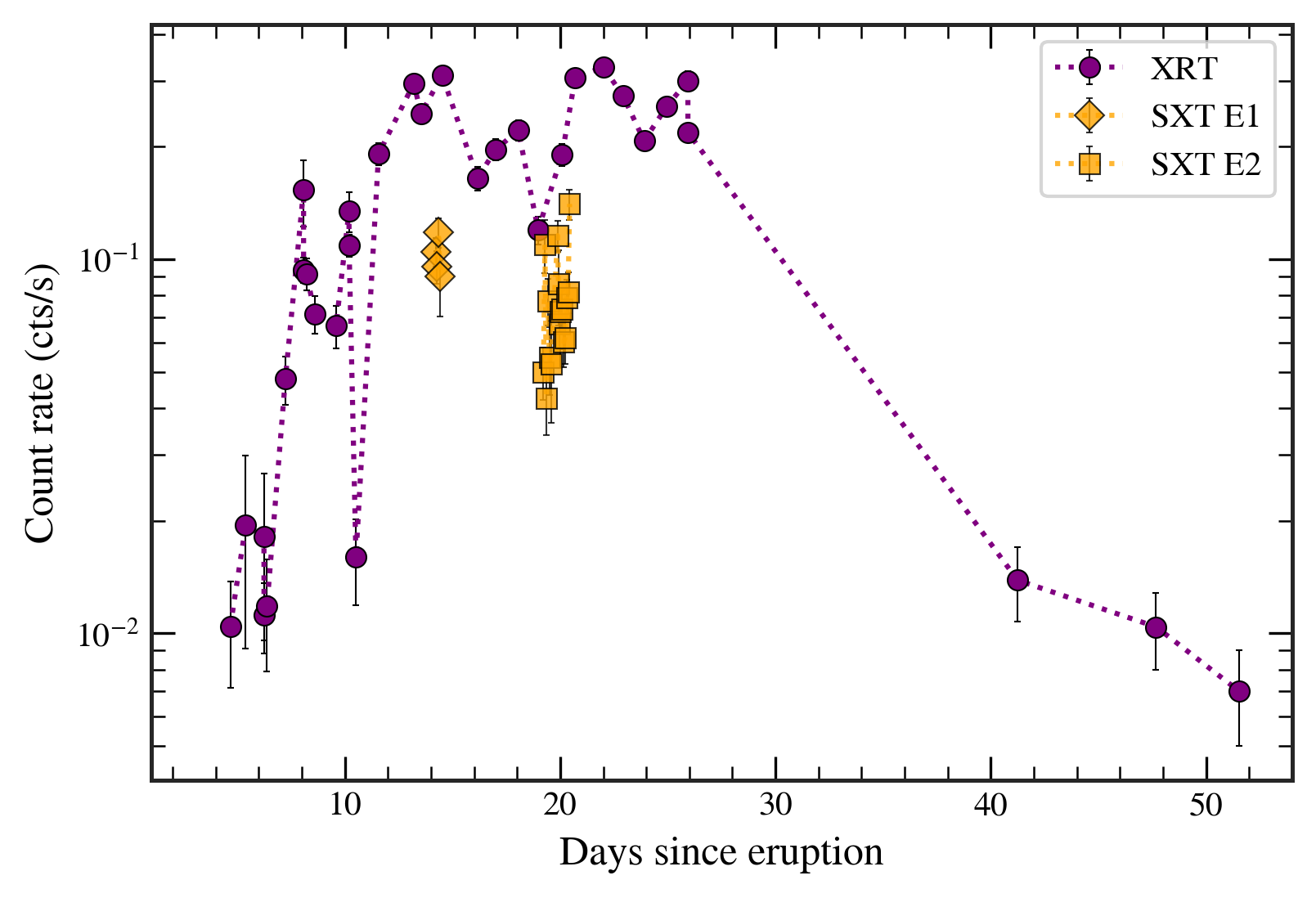}
    \includegraphics[width=0.49\linewidth]{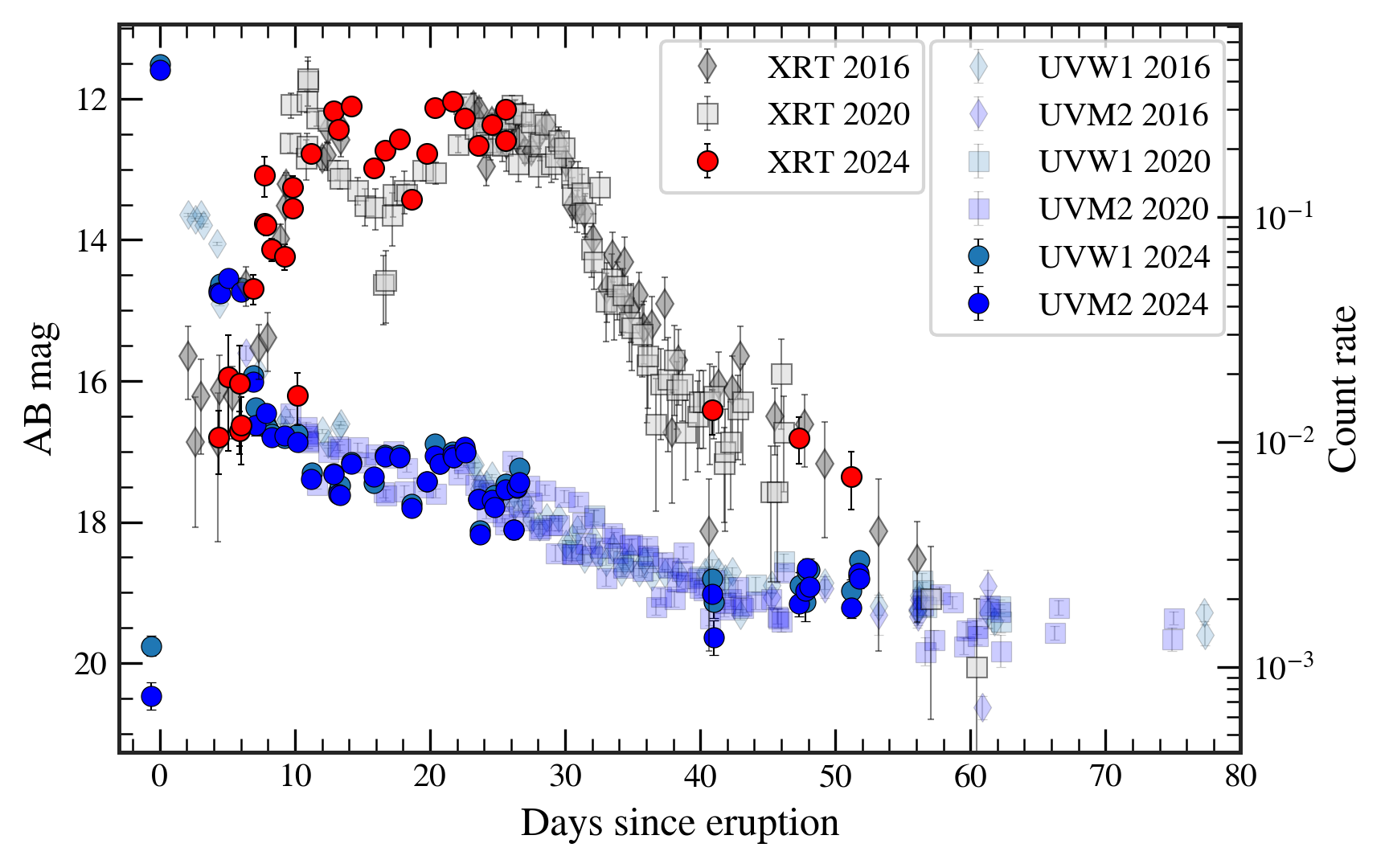}
    \includegraphics[width=0.49\linewidth]{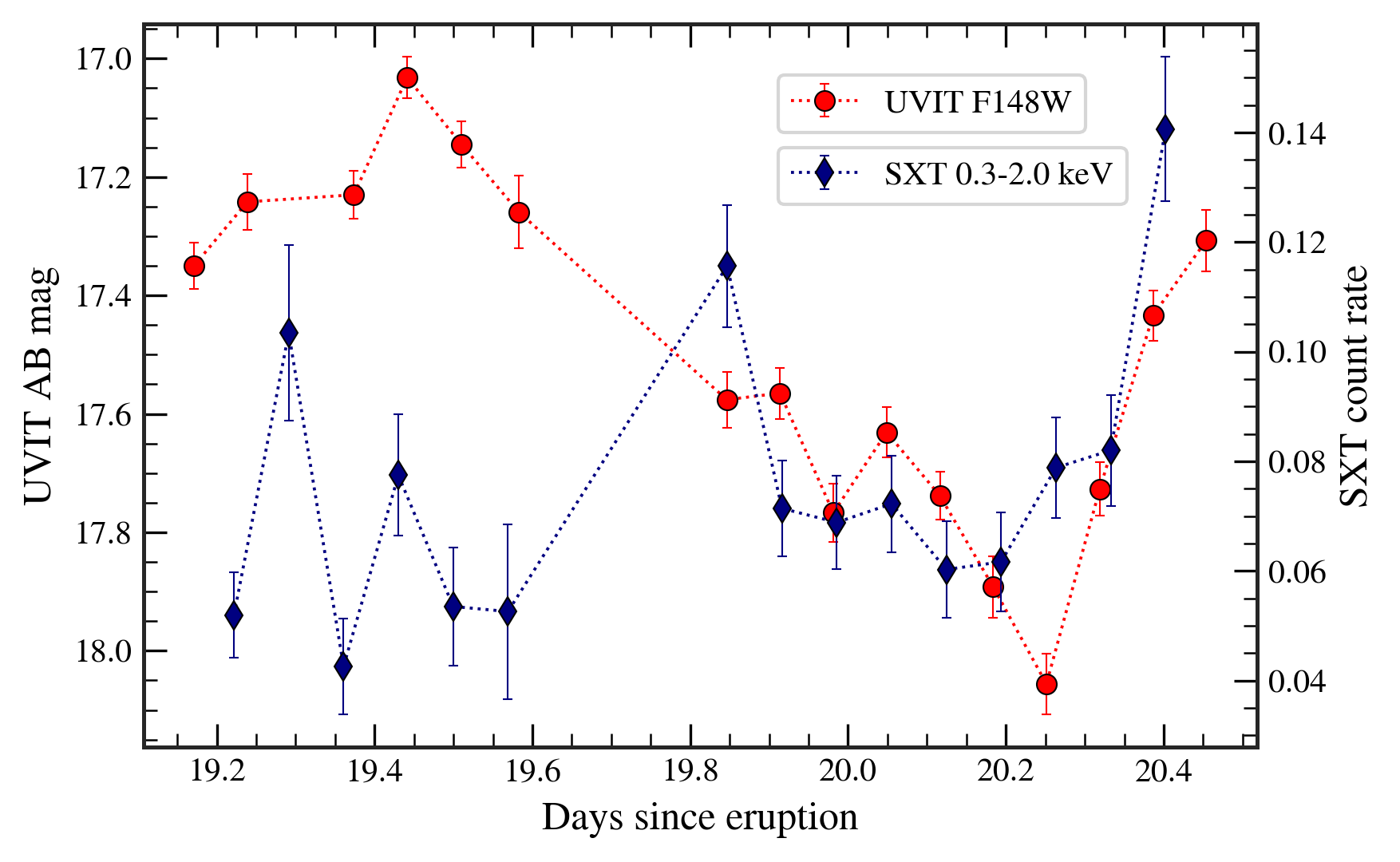}
    \caption{UV and X-ray light curves of LMCN 1968-12a. \textit{Top:} UV (left) and X-ray (right) light curves of the 2024 eruption of LMCN~1968-12a. \textit{Bottom left:} UV and X-ray light curves of the 2024 eruption overplotted on the light curves of previous eruptions. \textit{Bottom right:} Zoomed-in view of the second epoch of \textit{AstroSat} observations, showing variability consistent with a 1.26-day period, though only a single cycle is covered. The 2024 eruption data are available in machine readable format.}
    \label{fig:lightcurves}
\end{figure*}

\cite{Darnley_2024} reported the 2024 eruption of the RN on 2024 August 1, when it was observed to brighten by more than 8 mags in $\sim$ 15 hrs during the monthly monitoring of the RN field by \textit{Swift}. Soon, it was followed up in UV and X-ray wavebands by both \textit{AstroSat} and \textit{Swift} as discussed below. 

\subsection{AstroSat}
\textit{AstroSat} \citep{KPS_2014} is a space-based observatory carrying three imaging telescopes: two Ultraviolet Imaging Telescope (UVIT), one for NUV and the other for FUV channel \citep{Tandon_2017}, and a Soft X-ray Telescope (SXT; \citealt{KPS_2017}). Each UVIT telescope has a primary diameter of 37.5~cm with a field of view of 28\arcmin. Both NUV and FUV have multiple filters \citep{Tandon_2017}. We used only the FUV channel with F148W and F172M filters. The NUV channel was not available. UVIT has a spatial resolution of around 1.5\arcsec. In soft X-rays, SXT can observe in the 0.3-7~keV range with a spectral resolution of about 90 to 130 eV and a spatial resolution of 4\arcmin. We utilized channels in the 0.3-2 keV range, which is suitable for studying the supersoft source phase of novae \citep{Bhattacharya_2021}. 

After the discovery, we submitted a Target of Opportunity (ToO) proposal for \textit{AstroSat} UV and soft X-ray observations of LMCN 1968-12a. The first set of observations was performed on 2024 August 15, fourteen days after the outburst (Epoch I). It involved two orbits of UVIT observations: one using the $F148W$ filter with an exposure of around 180~s and another using the $F172M$ filter with an exposure of around 760~s. Simultaneously, SXT observed the object with a total effective exposure time of around 6600~s. A second ToO proposal was submitted to obtain a deep exposure during the supersoft phase, resulting in \textit{AstroSat} observations of the object again on 2024 August 20-21 (Epoch II). It included 16 orbits of UVIT observations, amounting to 12.6~ks of useful exposure with the F148W filter. The corresponding useful SXT exposure time was 23.2~ks, from all orbits. 

The UVIT L1 data were downloaded from the \textit{AstroSat} archive and processed using the \texttt{CCDLAB} software \citep{Postma_2021} to obtain the L2 data. Adhering to the steps in \cite{basu_2024a}, PSF photometry was performed, followed by aperture correction derived from bright stars and zero-point corrections taken from \cite{Tandon_2020}. For SED analysis, UV magnitudes were de-reddened by $E(B - V) = 0.07 \pm 0.01$, taken from the previous study \citep{Kuin_2020}.

Orbit-wise SXT data were downloaded from the \textit{AstroSat} archive and merged together with the \texttt{SXTMerger} package in \texttt{JULIA}. The SXT data were binned at 3000 s to generate light curves with an optimum signal-to-noise ratio (SNR) using \texttt{xselect (v2.4m)} in \texttt{HEASoft}. SXT spectra were extracted from orbit-wise, cleaned event files using \texttt{xspec (v12.12.0)}.  A circular region of 12\arcmin~ and an annulus of 15\arcmin~ to 17\arcmin~ radius were chosen as the source and background region, respectively, to extract the counts. The latest available Ancillary Response Files (ARF) and Response Matrix Files (RMF) were used for the analysis of spectra. The spectral analysis involved Poisson statistics (cstat), interstellar medium (ISM) abundances given in \cite{Wilms_2000}, and the T\"ubingen-Boulder (\texttt{tbabs}) ISM absorption model to account for neutral H in the intervening medium. Here, we assume $\mathrm{N_H}=1.8\times10^{21}$ cm$^{-2}$, which is the column density of the ISM and has been shown to be undisturbed by the optical depth of the ejecta after 10 days from eruption by \cite{Kuin_2020}.

The UV and X-ray light curves are shown in Figure~\ref{fig:lightcurves}, and some of the X-ray spectra are shown in Figure~\ref{fig:xray_spec}.

\subsection{Swift}
\textit{Swift} \citep{Gehrels_2004} consists of the UV/optical telescope (UVOT; \citealt{Roming_2005}) and the X-ray telescope (XRT; \citealt{Burrows_2005}). The UVOT has a spatial resolution of 2.3\arcsec, whereas XRT has a spatial resolution of 18\arcsec and a spectral resolution similar to that of SXT. \textit{Swift} monitored the object every month to look for any new outbursts. As soon as the 2024 outburst was detected \citep{Darnley_2024}, \textit{Swift} started daily monitoring from August 1, 2024, which lasted up to August 28 with both UVOT and XRT. It pointed to the object again on September 11, 18, and 22 after a gap of almost two weeks due to observational constraints. Publicly available L2 data were downloaded from the \textit{Swift} archives. 

UVOT data were analysed using the \texttt{uvotmaghist (v1.3)} command in \texttt{HEASoft}. A source region of 5\arcsec~centered at the source was used for aperture photometry, and a similar aperture in a source-free region near the nova was used to estimate the background. UVOT observations included $V, B, U, UVW1, UVM2,$ and $UVW2$ filters. Observations using $B$ and $V$ filters are restricted to the first 10 days, whereas complete light curves are available for the other filters, shown in Figure~\ref{fig:lightcurves}. 

XRT data were analysed using the \texttt{ximage/sosta (v4.5.1)} package within \texttt{HEASoft}, which corrects the counts for vignetting, dead time loss, background subtraction, and the PSF of the instrument. ARF was generated from the exposure maps to extract the spectra, while RMF was taken from the calibration database. Other parameters used in \texttt{xspec} were kept the same as for the SXT spectra analysis. XRT light curve and a couple of XRT spectra are shown in Figure~\ref{fig:lightcurves} and Figure~\ref{fig:xray_spec}, respectively. The temperature and $\mathrm{N_H}$ were also independently determined from the XRT spectra and their evolution is displayed in Figure~\ref{fig:xray_spec_param_evol}. The \texttt{error} command was used to determine the 68\% and 90\% confidence intervals for these parameters.  $\mathrm{N_H}$ at $>$ 12 days from eruption was found to be largely consistent with the previous results obtained by \cite{Kuin_2020}.

The \textit{AstroSat} and \textit{Swift} data are available in the form of MRTs.

\section{Results}
\label{section:results}
\subsection{Light curves}
\label{sec:light curves}

\textit{Day 0 to 8:} LMCN 1968-12a reached its maximum magnitude of 9.9 $\pm$ 0.1 on 2024 August 1.83 in the $UVM2$ filter \citep{Darnley_2024}. The UV and optical light curves follow a steep decline up to day 8 from the peak of the eruption, as shown in Figure~\ref{fig:lightcurves}. The rapid decline rate during this phase is $\sim$ 0.8~mags~day$^{-1}$ in all UVOT filters. The soft X-ray flux emerged during this UV-decline phase around day 5, and increased from the detection limit (0.014 counts s$^{-1}$) to more than 0.05 counts s$^{-1}$ in the XRT bands. 

\textit{Day 8 to 24:} Modulations with a period of $\sim$ 1.3 days are observed in all UV bands during this phase and are attributed to the orbital period \citep{Kuin_2020}. The high-cadence UVIT data around day 20 show evidence consistent with the 1.26-day orbital period, although the observations span only a single cycle, as illustrated in the bottom right panel of Figure~\ref{fig:lightcurves}. \cite{Kuin_2020} observed that the eclipse duration is only 0.05 parts of a full phase. Some UVOT data points, such as those on days 19 and 24, show significant dips in the light curve. The UVIT $F148W$ flux also shows a dip between days 19.8 and 20.4 (\textit{AstroSat} Epoch II).

The peaks of these light curve modulations in the UV bands show a gradual decline of 1 mag in 16 days (see the top-left panel in Figure~\ref{fig:lightcurves}), representing the plateau phase, which is often seen in RNe \citep{Pagnotta_2014}. The UV light curve during the plateau phase is very similar to the one seen in U~Sco, a galactic RN with a similar orbital period and a slightly longer recurrence period \citep{Evans_2023, Muraoka_2024}.

Another common feature observed during the plateau phase is the rise to the maximum of super-soft source (SSS) emission, originating from the WD's surface. As shown in Figure~\ref{fig:lightcurves}, the XRT count rate increases and reaches a peak flux of 0.3~counts~s$^{-1}$ during this phase. The SXT count rates, known to be approximately 0.3 times the XRT count rates (owing to lower sensitivity), follow the XRT light curve. Similar to the UV light curve, the SSS light curve also shows orbital modulations. Additionally, the SSS light curve has a double-peaked structure with a decrease in the flux during days $\sim 16-20$. Similar features were seen in the previous outbursts, with the broad inflection being more prominent, as shown in the lower left panel of Figure~\ref{fig:lightcurves}. The Epoch II \textit{AstroSat}-SXT observations indicate a sharp dip during days 19.8 and 20.4 (bottom right plot in Figure~\ref{fig:lightcurves}), which is more akin to the orbital modulation.

\subsection{X-ray Spectroscopy}
\begin{figure*}
    \centering
    \includegraphics[width=\linewidth]{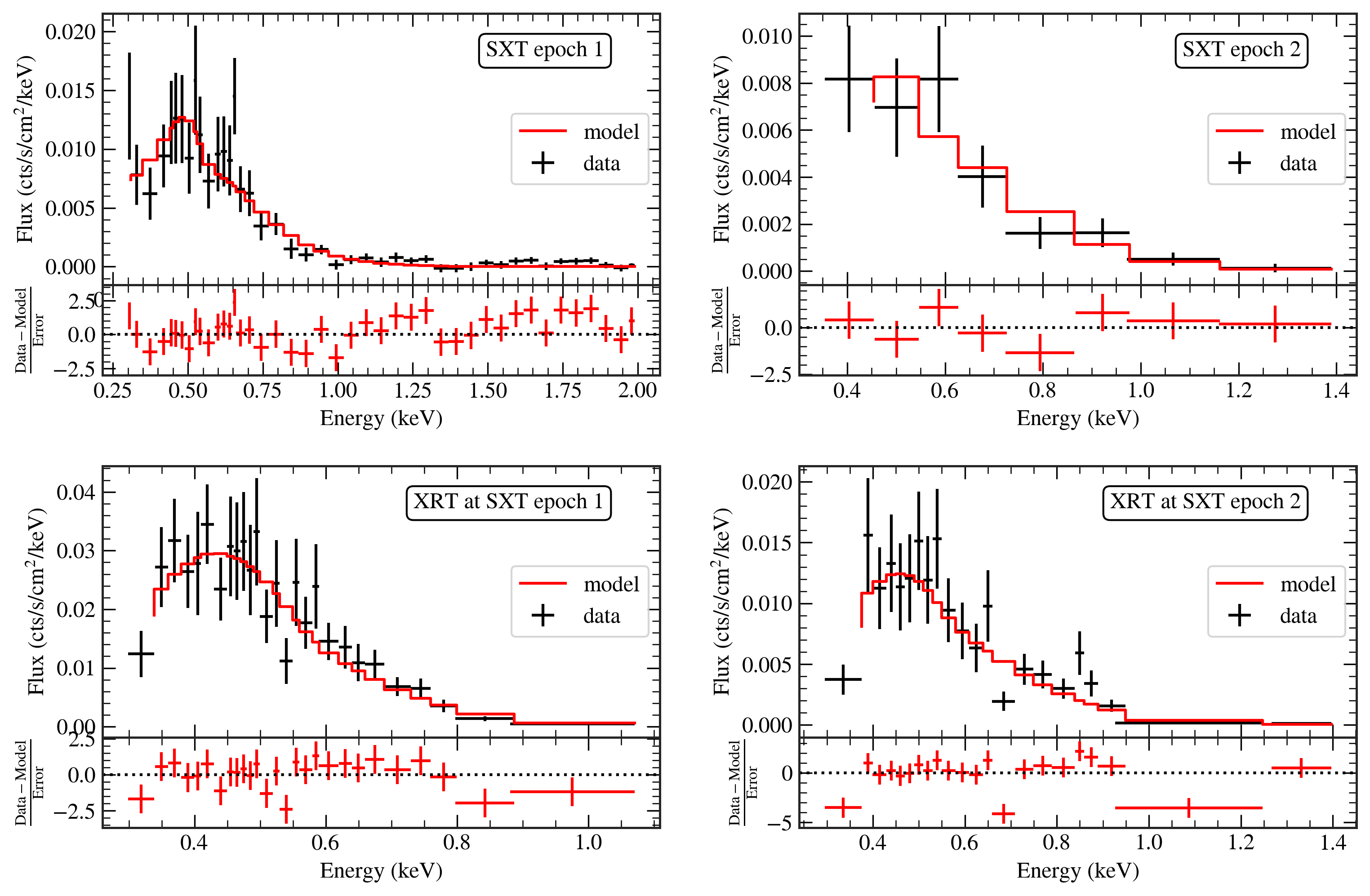}
    \caption{\textit{Top panel}: AstroSat/SXT spectra observed at Epoch I (2024 August 15 16:19-22:35 UT; $t_{\mathrm exp} \approx 100~$min) and Epoch II (2024 August 21 14:28-14:58 UT; $t_{\mathrm exp} \approx 30$ min). \textit{Bottom panel}: Swift/XRT spectra on 2024 August 16  00:23-00:45 UT ($t_{\mathrm exp} \approx 22$ mins) and August~21~14:14-14:36 UT ($t_{\mathrm exp} \approx 22$ mins), at epochs close to the \textit{AstroSat} observations. Best-fit \texttt{tbabs~$\times$~bb} models are shown in red. The fit residuals are also shown for each subplot. The best-fit parameters are given in Figure~\ref{fig:SED}. }
    \label{fig:xray_spec}
\end{figure*}

SXT spectra were obtained during both Epoch I (around day 15) and Epoch II (around day 20). For Epoch I, spectra from all four orbits were combined, as their count rates were consistent within the uncertainties, and the merged spectrum is shown in Figure~\ref{fig:xray_spec}. In contrast, the deeper Epoch II exhibits significant variability (see Figure~\ref{fig:lightcurves}). Therefore, only the spectrum from the orbit closest in time to an XRT observation is presented in Figure~\ref{fig:xray_spec}. From the light curve in Figure~\ref{fig:lightcurves}, it is evident that both epochs occurred during the peak of the supersoft phase. In Epoch I, the spectral evolution suggests that most of the radiation is below 1 keV, owing to its supersoft nature. During Epoch II as well, most of the counts are concentrated below 1 keV, with almost no photons above 1.5 keV. The dense temporal evolution during the second epoch shows highly variable spectra during the supersoft phase. The flux density rises almost five times and declines back to the initial level in 4-5 hours during this epoch of observation. Towards the end of the observation of Epoch II, the 0.3-1.0 keV flux starts to rise again. These variabilities are also evident from the X-ray light curves shown in Figure~\ref{fig:lightcurves}. The high variability of SSS emission below 1~keV lasting over timescales of hours is discussed in \S \ref{subsec:discussion_lc}. Although the spectral evolution hints at transient features consistent with lines seen during the SSS phase in novae \citep{Orio_2020, Ness_2022}, the limited resolution of the X-ray data in this case limits any noteworthy interpretation.

Nonetheless, these X-ray spectra are still useful for deriving physical parameters of the source region, close to the surface of the WD. The X-ray spectra from the two individual SXT epochs and their corresponding nearest XRT spectra are shown in Figure~\ref{fig:xray_spec}. These X-ray spectra were modeled with blackbody functions for their simplicity and reproducibility. It should be noted that blackbody fits are not physically accurate representations of the WD's atmosphere. But, given the limited spectral resolution of the data, it still provides a reasonable approximation of key parameters such as the effective temperature of the surface of the WD photosphere. The observed data points together with the corresponding best-fit models and the residuals are also shown in Figure~\ref{fig:xray_spec}. For the \textit{AstroSat}-SXT data, the best-fit temperatures during Epochs I and II are of order $\sim10^{6}$ K. It is consistent with the near-simultaneous Swift/XRT measurements as shown in Figure~\ref{fig:SED}. Extending this analysis to all the XRT data due to its good cadence throughout the SSS phase, we could obtain the temporal change of the effective temperature and the neutral hydrogen column density. This evolution is shown in Figure~\ref{fig:xray_spec_param_evol}. We notice that the rise and drop in X-ray flux corresponds to similar behaviour in the temperature as well. The $N_H$, on the other hand, stabilises around $1.5\times 10^{21}$ cm$^{-2}$ after day 13 which was also reported by \cite{Kuin_2020} for the 2016 eruption. It is consistent with the ISM $N_H$ values derived from other sources. However, it is interesting to note that $N_H$ shows an increase in a couple of instances corresponding with a drop in the flux and temperature. This behaviour has been explored in \S \ref{subsec:discussion_lc}.

\section{Discussions}

\subsection{The double-peaked nature of X-ray light curve}
\label{subsec:discussion_lc}

\begin{figure}
    \centering
    \includegraphics[width=\linewidth]{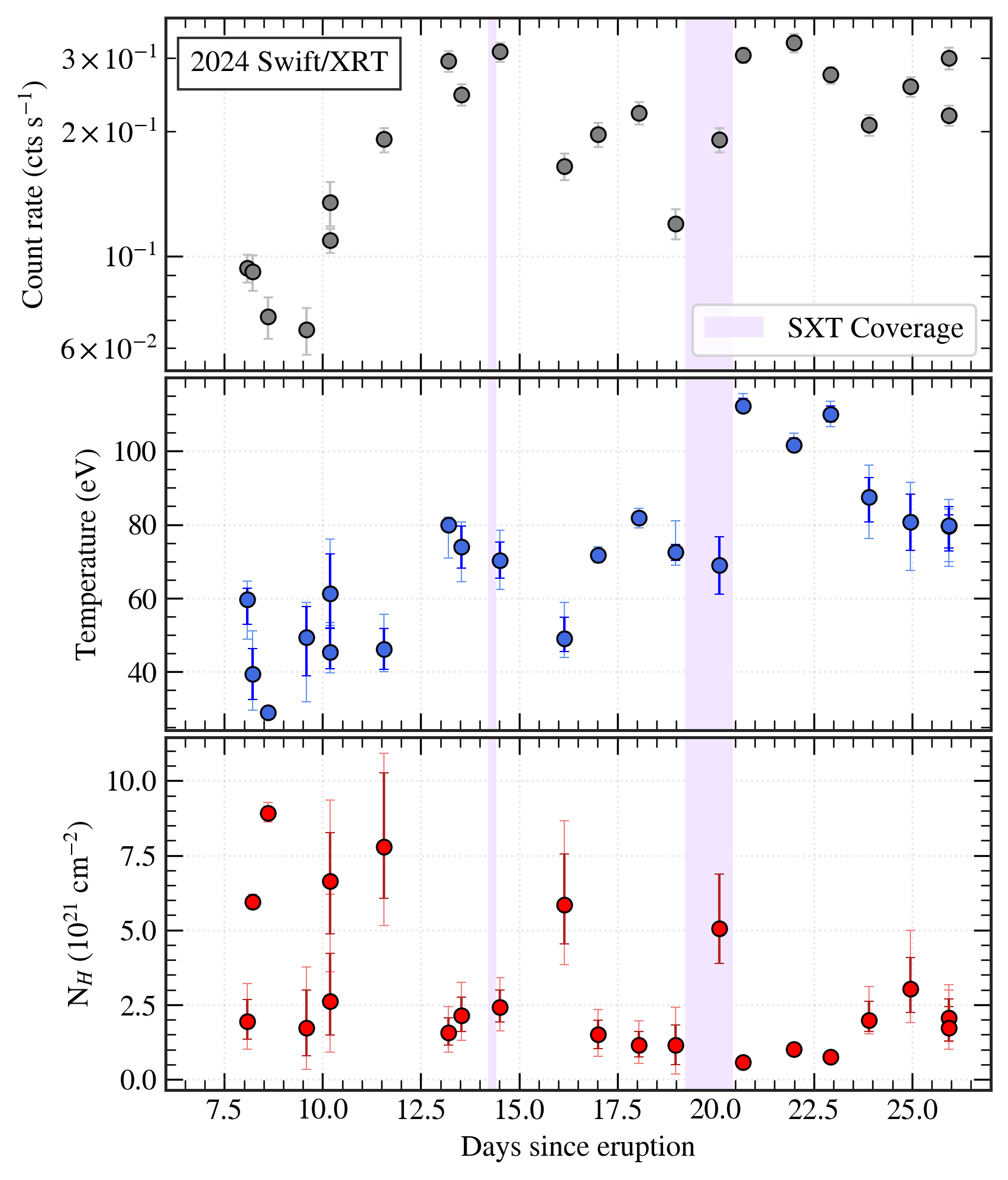}
    \caption{XRT light curve of the 2024 eruption is shown in the top panel. Evolution of the supersoft black body temperature and N$_{H}$ are plotted in the middle and bottom panels, respectively. The 68\% and 90\% confidence intervals for temperature and N$_H$ are also shown.}
    \label{fig:xray_spec_param_evol}
\end{figure}

The rise of the SSS emission is due to the ejecta becoming optically thin while expanding \citep{Krautter_1996}. This is evident from the drop in the column density noticed in Figure~\ref{fig:xray_spec_param_evol}. The expanding ejecta reveals the X-ray photons originating from deeper layers near the surface of the WD. The origin of UV photons, however, is debated.

It was noted in \S \ref{sec:light curves} that on days 19 and 24, the UV flux drops sharply in all the UVOT filters. By overplotting the eclipse durations on the flux evolution in Figure~\ref{fig:BB_fit_evol}, we notice that these dips are actually caused by the eclipsing of the WD by the secondary. Since the UV light curve shows eclipses, it must have been emitted from somewhere within the orbit of the secondary. The simultaneous drop in UV--X-ray flux during day 19.8 to 20.4 noticed in the high cadence \textit{AstroSat} Epoch II observations (bottom right plot in Figure~\ref{fig:lightcurves}) must also be because of the secondary star eclipsing the WD and the disk. As the WD emerges from behind the secondary, the UV--X-ray fluxes start to rise again after day 20.6. However, this dip around day 20 is not related to the overall wide drop in the SSS light curve observed between days 16--20. The timescale and intensity of the broader drop, revealing the light curve as a double-humped peak, differ from those attributed to orbital modulation. Consequently, it is reasonable to assume that orbital motion is not a cause of the double humped SSS peak.

Double-humped SSS emission, along with high variability, has been observed in other well-studied RNe. V3890~Sgr displayed a highly variable X-ray light curve, featuring dips on time scales ranging from hours to days \citep{KPS_2021, Ness_2022}. The hourly dips were attributed by \citet{Ness_2022} to clumps of ejecta orbiting within the system at radii of 5–150 $\mathrm{R_{WD}}$. In contrast, an extended low-flux interval from day 16.8 to 17.8, or a ``faint X-ray episode", was probably caused by a sudden outflow of matter \citep{KPS_2021}. This outflow was hypothesised to be partially opaque to X-rays but with a low filling factor, allowing the observed spectrum and column density ($N_H$) to remain largely unaffected, though $N_H$ estimation was limited by low signal-to-noise ratio. In particular, this episode lacked the short-term variability seen at other times.

RS~Oph exhibited diverse variability during its 2006 and 2021 eruptions. \citet{Osborne_2011} reported high-amplitude variations on time scales of tens of hours during the rise of the SSS phase in 2006, attributed to variable absorption by clumpy ejecta. In comparison, the 2021 SSS light curve was fainter and displayed a multi-peaked structure. \citet{Ness_2023} proposed that these differences arise from varying optical depths due to inhomogeneous ejecta across different outbursts. Interestingly, the 2021 eruption also featured a broad dip in X-ray emission between days 47 and 57 (see Figure~1 in \citealt{Ness_2023}), similar in duration to that seen in LMCN~1968-12a, though its origin remains unclear.

The soft X-ray light curve of U~Sco during its 2010 eruption showed an initial rise followed by a plateau phase of nearly constant flux until day 20, before reaching its peak around day 30 \citep{Pagnotta_2015}. Until day 25, it also exhibited, sharp aperiodic dips lasting several hours, attributed to occultations of the central source by dense absorbing material aligned with the trajectory of a reforming accretion stream \citep{Ness_2012}. For the 2022 eruption, \citet{Muraoka_2024} reported that the early X-ray emission was fainter than in 2010, probably due to a lower accreted mass and consequently a reduced post-eruption envelope mass.

M31N~2008-12a has also shown high variability in X-rays \citep{Darnley_2016, Henze_2018}. Like LMCN~1968-12a, it also demonstrates a major dip near the midpoint of the SSS phase, consistently over almost all eruptions lasting more than a day (see Figure~7 in \citealt{basu_2024}). 

Short-timescale variability, lasting from several minutes to a few hours, is commonly attributed to clumpy ejecta or inhomogeneous structures such as reforming accretion streams, which obscure the WD photosphere. The depth and duration of these dips are expected to vary between eruptions. However, in systems like LMCN~1968-12a and M31N~2008-12a, the observed day-scale dips appear to be remarkably consistent across multiple eruptions. In particular, the double-peaked structure in LMCN~1968-12a X-ray light curve (see bottom left panel of Figure~\ref{fig:lightcurves}) is repeatedly observed in at least its two most recent eruptions. We probed the blackbody effective temperature (kT) and the neutral hydrogen column density (N$_\mathrm{H}$) from the XRT data of the supersoft phase in 2024 by fitting a simple \texttt{tbabs~$\times$~bb} model, keeping all parameters free. As shown in Figure~\ref{fig:xray_spec_param_evol}, the drop in flux between days 16 and 20 indicates a correlated drop in temperature. An interesting point to note is that N$_\mathrm{H}$ increases from the ISM value during the days corresponding to the decrease in the X-ray flux and temperature. Since the dip from day 16 to 20 is much wider than the usual faster time-scale variability seen in the SSS phases of recurrent novae, and occurs at the same phase during the two outbursts, it is unlikely to be caused by clumps in the ejecta. It is more likely to be caused by obscuration due to external inhomogeneous material. 

Recent studies have found faint nebulosities around recurrent novae M31N~2008-12a \citep{Darnley_2019}, KT~Eri \citep{Shara_2024, Kalesh_2024a}, RS~Oph \citep{Kalesh_2024, Shara_2025}, and T~CrB \citep{Shara_2024a} that are orders of magnitude larger than the usual nova shells. These substantially extended nebulosities, termed nova super-remnants \citep{Kalesh_2023}, are believed to be a result of ISM swept up by repeated nova eruptions. Such structures are expected to be more prominent in the case of RNe due to their multiple nova outbursts over thousands of years. The presence of such an extended, inhomogeneous structure along the line-of-sight could possibly explain the dips lasting for days seen during the SSS phase in LMCN~1968-12a, and possibly M31N~2008-12a.

\subsection{Signatures of the surviving accretion disk}

\begin{figure*}
    \centering
    \includegraphics[width=.49\linewidth]{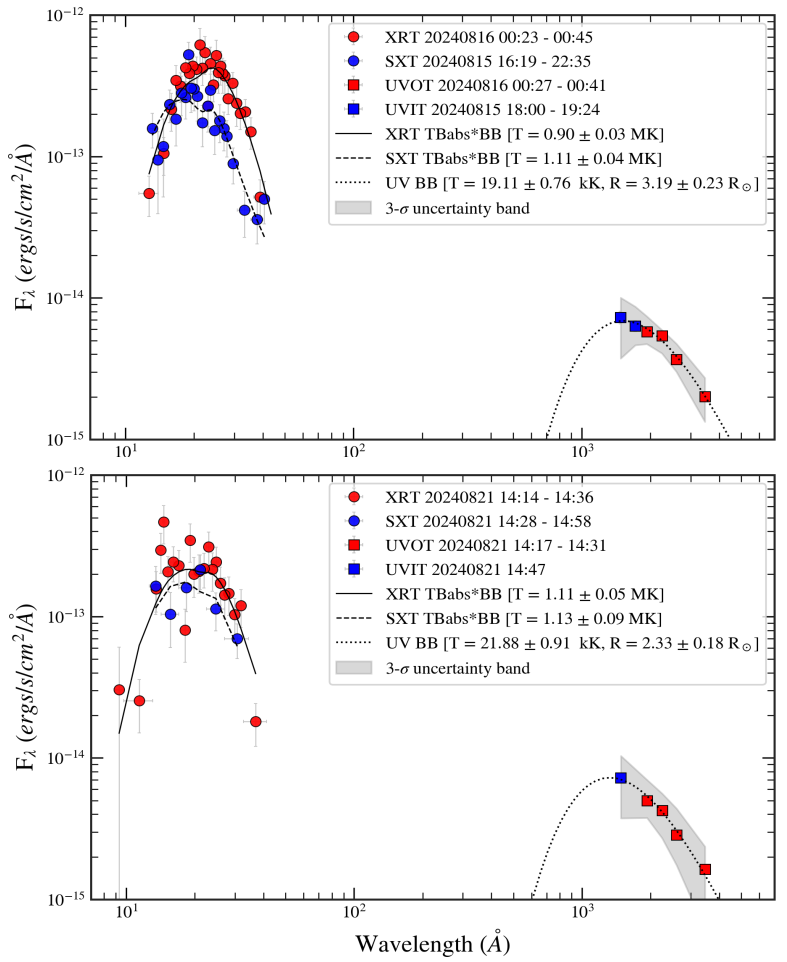}
    \includegraphics[width=.50\linewidth]{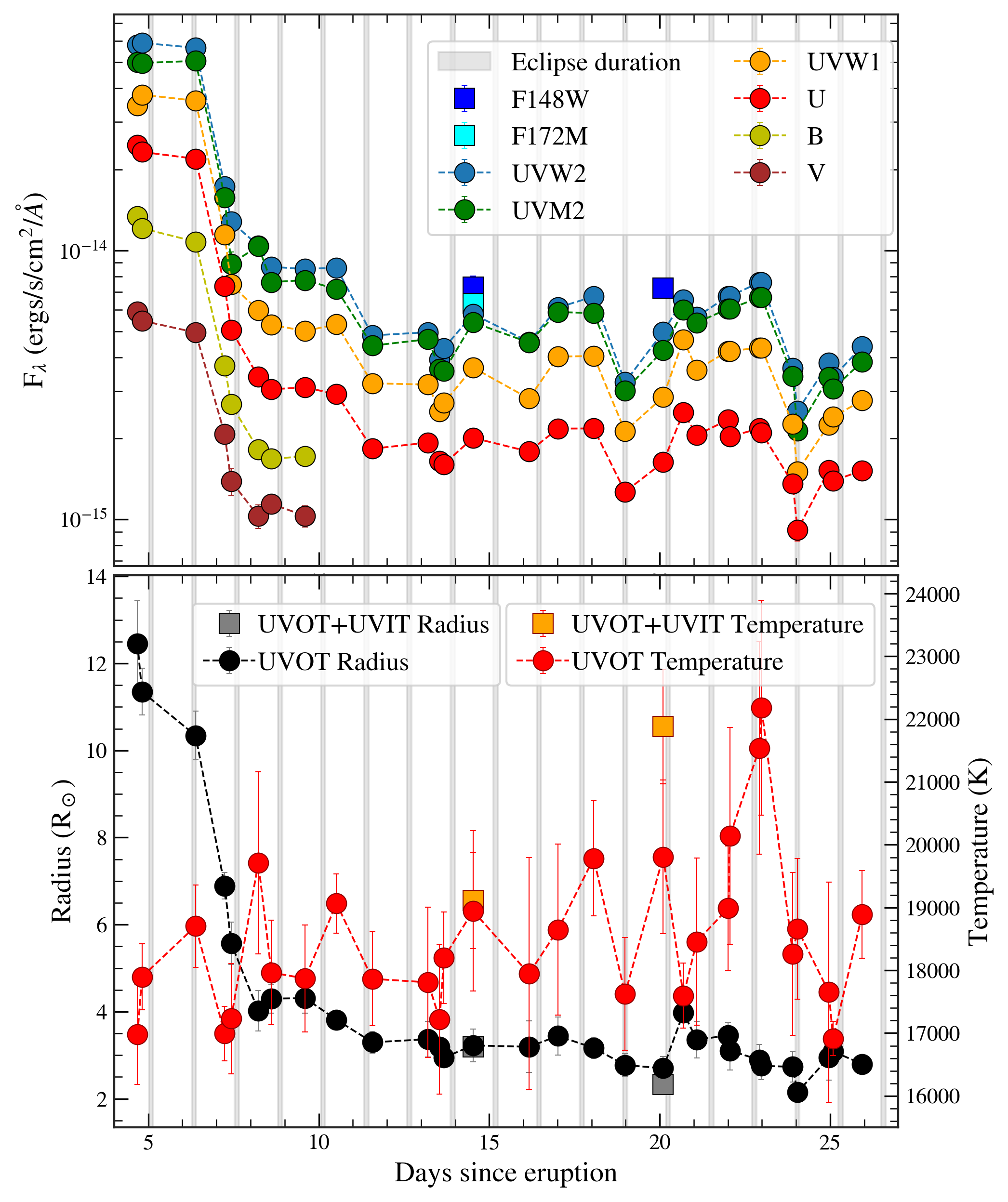}
    \caption{\textit{Left:} Blackbody fits to the X-ray (SXT and XRT) and UV (UVIT and UVOT) data are overplotted on the observed spectra for epoch I (top) and epoch II (bottom); the corresponding best-fit parameters are listed in the legends. \textit{Right:} The top panel shows the temporal evolution of the flux in different UV filters, while the bottom panel presents the evolution of the inferred model parameters, i.e., the radius and effective temperature of the UV-emitting source.}
    \label{fig:BB_fit_evol}
    \label{fig:SED}
\end{figure*}

Simultaneous NUV, FUV, and X-ray observations with \textit{AstroSat} and \textit{Swift} motivated us to look into the spectral energy distribution (SED) of the evolving system. SEDs covering the X-ray--UV wavebands during \textit{AstroSat} Epochs I and II are shown in Figure~\ref{fig:BB_fit_evol}. It is evident that the soft X-ray and UV fluxes originate from different sources. A single blackbody fit to the combined X-ray and UV data did not yield a good fit, with the X-ray tail being much weaker in the UV wavebands than what the observations suggest. As a result, the X-ray and UV data were modeled separately. The X-ray component was fitted with a blackbody modified by an absorption column mentioned in \S \ref{section:obs}, whereas the extinction-corrected UV component was fitted only with a blackbody model. The results for \textit{AstroSat} Epochs I and II, combined with near-simultaneous \textit{Swift} data, are shown in Figure~\ref{fig:BB_fit_evol}. The X-ray component, which is responsible for the SSS phase, peaks at around $10^6$~K, which was also reported by \cite{Kuin_2020} and suggested to originate at the surface of the WD. Meanwhile, the UV component peaks at roughly $2\times 10^4$~K and has a source size of 2-3~R$_{\odot}$ during these epochs (day $\sim$15 and $\sim$20). 

The \textit{Swift} UVOT data were also individually fitted with blackbodies to estimate the temporal variation in radius and temperature of the source as shown in Figure~\ref{fig:BB_fit_evol} (bottom-right panel). The radius of the UV source decreases till day 8, which coincides with the beginning of the SSS phase. This decrease in radius from 12~R$_{\odot}$ to 4~R$_{\odot}$ traces the receding photosphere towards the WD surface, which subsequently gives rise to the SSS phase in X-rays and a plateau phase in UV-optical bands. During the SSS phase, the radius of the UV source stays nearly constant at roughly 3~R$_{\odot}$, while the temperature is estimated to be between 1.7-2.3~$\times 10^4$~K, consistent with the results from \cite{Evans_2025}. The UV blackbody component cannot originate from the secondary evolved star or the accretion disk alone. However, if any or both of these are irradiated by the hot WD, having a temperature that peaks at ~$10^6$~K, it can give rise to UV-optical blackbody radiation. \cite{Kuin_2020} and \cite{Evans_2025} ruled out the possibility of an irradiated secondary as the model predictions in such a case overestimate the observed UV-optical fluxes. 

The survival of the accretion disk after a nova outburst has been debated over the years. Recent studies of rapid recurrent novae such as M31N~2008-12a \citep{basu_2024} and U~Sco \citep{Ness_2012,Muraoka_2025} indicate that the disk is present during the supersoft phase, even if in a fragmented form due to the impact of the nova eruption. U~Sco showed signatures of accretion stream as early as 8 days after eruption \citep{GCA_2013}. Recent simulations by \cite{Figueira_2025} also support the disk survival scenario under certain circumstances for RNe like U~Sco. Disk irradiance is not rare and has been suggested to be the case for the optical plateau in U~Sco \citep{Muraoka_2024}, an RN with a similar orbital period and the same order of recurrence period as LMCN~1968-12a. \cite{Hachisu_2000} generated theoretical models of an irradiated disk and secondary for U Sco, which could reproduce its UV-optical light curve. It is more than likely that the same phenomenon is also responsible for LMCN~1968-12a, given the similarities between these two RNe.

The accretion disk therefore appears to survive the nova outburst in this case. RNe typically host very massive WDs, which require a smaller ignition mass to trigger an eruption. This, in turn, leads to a relatively low ejecta mass \citep{Kemp_2024}. Under such conditions, the impact on the accretion disk is expected to be less severe than in CNe, increasing the likelihood of disk survival after eruption. A direct consequence is the rapid resumption of accretion in RNe, within timescales of the order of SSS $t_{on}$, which in turn alleviates their recurrent nature.

\section{Conclusions}
\label{section:conclusion}
In this work, we have analysed the UV-X-ray light curves along with X-ray spectra and UV SEDs of the 2024 eruption of LMCN 1968-12a, a nearby rapidly recurring nova with a massive WD. The UV light curves show an initial steep decline followed by a plateau phase, which is modulated by the orbital period. As the UV light curve declines, the photosphere recedes, and the X-ray emission becomes visible. During the SSS phase, the X-ray emission is variable and also shows a major dip across outbursts, which could be due to a large absorbing medium along the line-of-sight, such as a nova super-remnant. The X-ray spectra and the UV SED modeling indicate that the X-ray and UV emission come from different regions. Soft X-rays are emitted by residual burning on the WD surface at a temperature of $\sim 10^6$~K. At the same time, UV-optical emission is dominated by the irradiated accretion disk at a temperature of $2\times10^4$~K. It is therefore evident from this study and previous studies of rapidly recurring nova systems that the accretion disk is not fully disrupted and accretion resumes within days after eruption. 

\section*{Acknowledgement}

\begin{acknowledgments}
We would like to thank the anonymous referee for the valuable comments that improved the quality of this work. 

This work uses the SXT and UVIT data from the \textit{AstroSat} mission of the Indian Space Research Organisation (ISRO). We thank the \textit{AstroSat} TAC for allowing us ToO time to observe this nova in 2024. We thank the SXT and UVIT payload operation centers for verifying and releasing the data via the ISSDC data archive and providing the necessary software tools. 

We acknowledge the use of public data from the \textit{Swift} data archive. This work has also used software and/or web tools obtained from NASA's High Energy Astrophysics Science Archive Research Center (HEASARC), a service of the Goddard Space Flight Center and the Smithsonian Astrophysical Observatory. 

K.P.S. and G.C.A. thank the Indian National Science Academy (INSA) for support under the INSA Senior Scientist Programme.

\end{acknowledgments}

\vspace{5mm}
\facilities{\textit{AstroSat} UVIT and SXT, \textit{Swift} UVOT and XRT}

\software{\texttt{CCDLAB} \citep{Postma_2021}, 
         \texttt{IRAF v2.16.1} \citep{Tody_1993},
         \texttt{HEASoft v6.29}, \texttt{ximage v4.5.1}, and \texttt{xselect v2.4m} \citep{Heasarc_2014}, 
         \texttt{xspec v12.12.0} \citep{Arnaud_1996},
         \texttt{Python v3.8.17} \citep{python09}, 
         \texttt{NumPy v1.24.4} \citep{Harris_2020},
         \texttt{pandas v1.5.3} \citep{pandas_2010},
         \texttt{matplotlib v3.7.1} \citep{Hunter_2007},
         \texttt{astropy v5.0.6} \citep{astropy:2022}
         }



\newpage
\appendix

\bibliography{main}{}
\bibliographystyle{aasjournal}



\end{document}